\documentclass[prl,floatfix,twocolumn,showpacs,amsmath,amssymb]{revtex4}
\usepackage{graphicx}
\usepackage{dcolumn}
\usepackage{bm}

\begin{document}

\title{Exotic gapless spectrum induced by frustration in quantum 
antiferromagnets}
\author{Federico Becca,$^{1}$ Luca Capriotti,$^{2}$ Alberto Parola,$^{3}$ 
and Sandro Sorella$^{1}$}
\affiliation{
$^{1}$ CNR-INFM-Democritos National Simulation Centre and International School 
for Advanced Studies (SISSA), Via Beirut 2-4, I-34014 Trieste, Italy \\
$^{2}$ Global Modelling and Analytics Group, Investment Banking Division, Credit
Suisse Group, One Cabot Square, London, E14 4QJ, United Kingdom \\
$^{3}$ Dipartimento di Fisica e Matematica, Universit\`a dell'Insubria,
Via Valleggio 11, I-22100 Como, Italy
}
\date{\today}
\begin{abstract}
We show strong numerical evidence in favor of an unexpected virtually
gapless spectrum, with edge states localized at the boundaries, in 
frustrated spin-1/2 antiferromagnetic ladders with an odd number of legs.
These features can be accurately reproduced by using a projected BCS wave 
function with a non-trivial pairing, that mixes even and odd reflection
symmetries. This approach gives the correct classification of the excitations 
and provides a simple and very appealing picture of an unconventional 
spin-liquid phase stabilized by frustration.
\end{abstract}

\maketitle

In the last few decades there has been an increasing theoretical and
experimental effort to clarify the nature of the disordered phases 
stabilized by competing interactions in magnetic materials.~\cite{misguich}
A very exciting scenario appears when such disordered {\it spin liquids} 
are gapless with fractional excitations, generalizing the critical phase of 
one-dimensional systems to higher dimensions.~\cite{wen}
Recent realizations of frustrated antiferromagnets on quasi-two-dimensional
lattices, like ${\rm NiGa_2S_4}$ or 
${\rm \kappa{-}(ET)_2Cu_2(CN)_3}$,~\cite{nakatsuji,kanoda} give a promising 
evidence in this direction. In particular, in organic materials,
NMR shows a power-law behavior of $1/T_1$ at very low 
temperatures without any signal of magnetic order.~\cite{kanoda2}
Interestingly, also in ${\rm Cs_2CuCl_4}$, Neutron Scattering measurements 
show the existence of a continuum of excited states, compatible with pairs 
of spinon excitations.~\cite{coldea}

From a theoretical point of view, a 
promising strategy is to consider quasi-one-dimensional systems, where several
very reliable techniques are available, e.g., bosonization or 
density-matrix renormalization group (DMRG), in order to extract important 
insight into the more relevant two-dimensional (2D) case. In this Letter, we 
take this point of view and study the spin-1/2 $J_1{-}J_2$ Heisenberg model 
on ladders with an {\it odd} number of legs
\begin{equation}\label{hamiltonian}
{\cal H} = J_1 \sum_{\langle R,R^\prime \rangle} 
{\bf S}_R \cdot {\bf S}_{R^\prime}
+ J_2 \sum_{\langle \langle R,R^\prime \rangle \rangle} 
{\bf S}_R \cdot {\bf S}_{R^\prime},
\end{equation}
where ${\bf S}_R$ is the spin operator on site $R=(x,y)$, and the sum is 
restricted to first ($J_1$), second ($J_2$) nearest neighbors. 
We consider systems with $N=L \times n$ sites, where $n$ is the 
odd number of legs and $L$ is the number of rungs, with open boundary 
conditions along the rungs.  

The physical properties of odd-leg
ladders are expected to be similar to the ones of the 2D case. Indeed, 
although for any finite number of legs $n$ there are no magnetically ordered 
phases, in the weakly frustrated regime, i.e., $J_2/J_1 \ll 1$ or 
$J_1/J_2 \ll 1$, the excitation spectrum is gapless with a well defined 
spin-wave velocity, analogously to the 2D case.
Such a system, for small $n$, has been recently considered
by different groups with contradicting results.~\cite{nersesyan,balents,wang}
In particular, in Ref.~\cite{balents} it has been argued that, for any 
number of legs, a relevant operator that breaks the translational symmetry 
induces spontaneous dimerization in the intermediate region 
$J_2/J_1 \sim 1/2$. 
On the contrary, such a dimerization has not yet been detected 
by numerical calculations.~\cite{wang}
In order to clarify this issue, we consider the model of 
Eq.~(\ref{hamiltonian}) with $n=3$, by using Lanczos exact diagonalizations 
and DMRG. Moreover, by means of an accurate 
variational wave function we are able to interpret our numerical results.
 
\begin{figure}
\includegraphics[width=0.45\textwidth]{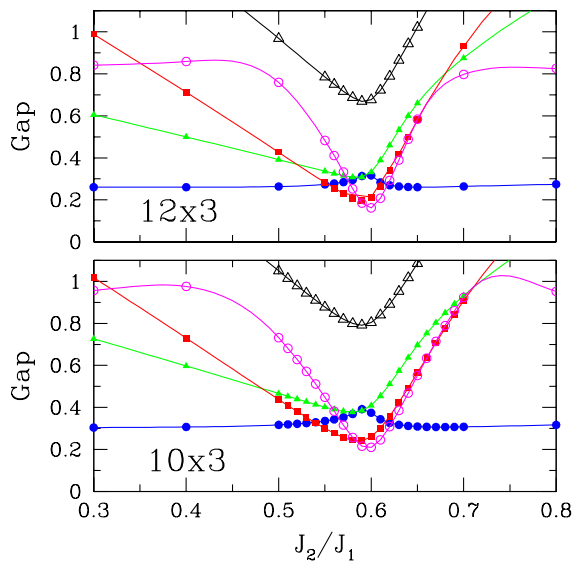}
\caption{\label{fig:spectrum}
(Color online) Some of the relevant low-energy gaps as a 
function of $J_2/J_1$ from Lanczos data. Triplet with $k_x=\pi$, $r_x=1$, 
and $r_y=1$ (blue circles), singlet with $k_x=\pi$, $r_x=-1$ and $r_y=1$ 
(green triangles), singlet with $k_x=0$, $r_x=1$, and $r_y=-1$ (red squares), 
singlet with $k_x=\pi$, $r_x=1$, and $r_y=-1$ (empty black triangles), 
and the lowest excitation in the same subspace of the ground state (empty 
magenta circles). All the quantum numbers are referred to the ground state 
and lines are guides to the eye.}
\end{figure}

Let us start with the excitation spectrum obtained by Lanczos diagonalizations 
with periodic boundary conditions along the chains.
In Fig.~\ref{fig:spectrum}, we show the evolution of few low-energy 
excitations in the relevant subspaces defined by the spatial symmetries of the
Hamiltonian: the crystal momentum
along the $x$ direction, the reflection ${\cal R}_x$ that changes $x \to -x$,
the reflection ${\cal R}_y$ across the central chain, and the total spin.
For both small and large $J_2$, as expected, all the excitations odd under 
${\cal R}_y$ (i.e., $r_y=-1$) have a sizable gap. 
Although the quantum numbers of the ground state
do not change by varying $J_2/J_1$, we have evidence in favor of an avoided
crossing, that is indicated by the sudden drop of the lowest energy gap in
the ground-state subspace. This suggests the possible occurrence of a 
first-order phase transition, slightly shifted to larger values of $J_2/J_1$
(i.e., $J_2/J_1 \sim 0.6$) with respect to the classical model.~\cite{misguich} 
More interestingly, the frustrating interaction induces a dramatic 
effect on excitations with odd ${\cal R}_y$, and for 
$J_2/J_1 \sim 0.55$ the lowest excitation is an $r_y=-1$ singlet with zero 
momentum $k_x$ (referenced to the ground state). 
Although Lanczos diagonalizations are limited to $12 \times 3$, we have 
evidence that, in the thermodynamic limit, this excitation becomes gapless 
in a small region around $J_2/J_1 \sim 0.55$. It should be emphasized 
that the quantum numbers of this excitation are not the ones implied
by the Affleck-Lieb-Schulz-Mattis theorem, i.e., a gapless state with 
momentum $k_x=\pi$ with respect to the ground state.~\cite{lsm} 
In agreement with that, we have also evidence that both a triplet state 
($r_x=1$ and $r_y=1$) and a singlet state ($r_x=-1$ and $r_y=1$) with $k_x=\pi$
become gapless in the thermodynamic limit for all the values of $J_2/J_1$ 
considered. These results are not compatible with conventional 
dimerization, where the ground state is doubly degenerate with a 
finite gap with respect to all the other excitations. 

\begin{figure}
\includegraphics[width=0.45\textwidth]{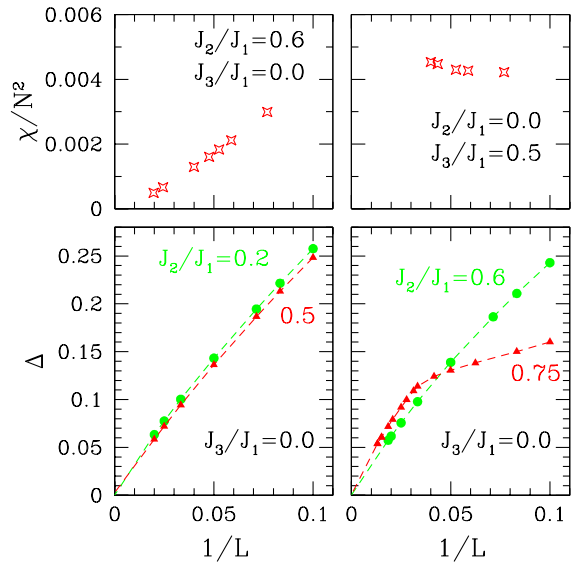}
\caption{\label{fig:scaldim}
(Color online) Size scaling for the dimer susceptibility $\chi$ (upper panels) 
and for the triplet gap $\Delta$ (lower panels) evaluated by DMRG.} 
\end{figure}

\begin{figure}
\includegraphics[width=0.45\textwidth]{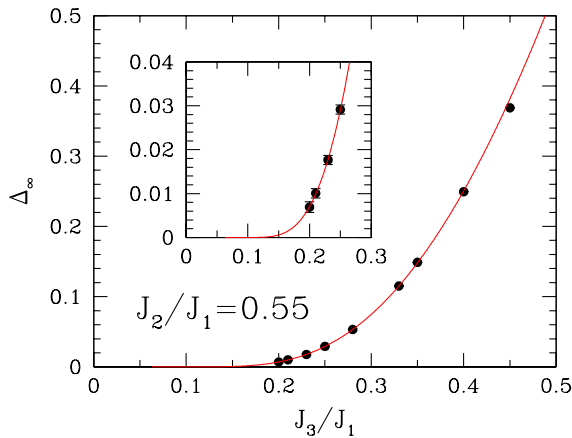}
\caption{\label{fig:j3crit}
(Color online) Triplet gap $\Delta_\infty$ as a function of $J_3/J_1$ in the 
thermodynamic limit, from DMRG data. The line is a fit according to the 
KT theory, $\Delta_\infty =A \exp[-b/\sqrt{(J_3-J_3^c)/J_1}]$. 
Inset: Detail for $J_3/J_1<0.3$.} 
\end{figure}

A further confirmation of a gapless triplet excitation comes from a systematic 
finite-size scaling by using DMRG with open boundary conditions in 
both $x$ and $y$ directions. The triplet gap vanishes in the thermodynamic 
limit for all the values of $J_2/J_1$ considered (see Fig.~\ref{fig:scaldim}), 
in agreement with previous DMRG results.~\cite{wang} In order to corroborate
this conclusion we have studied the influence of a 
third-nearest-neighbor interaction, $J_3$. As suggested
in Ref.~\onlinecite{caprio}, for $J_2=0$ this stabilizes a phase 
with a sizable gap and dimerization. In this respect, 
we have studied the size scaling of the spin gap for $J_2/J_1=0.55$ for 
different values of $J_3/J_1$. As shown in Fig.~\ref{fig:j3crit}, 
the thermodynamic limit of the spin gap is clearly finite for 
$J_3/J_1 \gtrsim 0.2$, and vanishes as $J_3$ decreases. Since the spin gap 
is characterized by a Kosterlitz-Thouless (KT) behavior, the precise value 
of the critical coupling $J_3^c$ is difficult to assess numerically. 
However, our results indicate that for $J_3/J_1 \lesssim 0.1$ the spin gap is 
exceedingly small ($\lesssim 10^{-8} J_1$). In addition, extensive 
calculations on five- and seven-leg ladders indicate that the finite-size gap
at fixed $L$ decreases with the number of legs, thus suggesting the
stabilization of a gapless and homogeneous phase for 
$J_2/J_1 \simeq 0.55$ also in 2D. This conclusion is also confirmed by a 
study of the dimer susceptibility calculated by adding a small 
perturbation that breaks the translational symmetry, 
$O = \delta \sum_{R} e^{i Q R} {\bf S}_R \cdot {\bf S}_{R+x}$, with 
$Q=(\pi,0)$, and by computing the second derivative of the ground-state energy 
with respect to $\delta$ (see Fig.~\ref{fig:scaldim}). A clear dimerization is 
found for $J_3/J_1 = 0.5$ but not for $J_3=0$, as in the latter case  
the dimer susceptibility does not display the
sharp $\chi \propto N^2$ divergence required in a symmetry broken
phase.~\cite{reviewcaprio} 

In the following, we will show that a simple variational ansatz 
is able to explain the gross features of the anomalous low-energy spectrum.
Our construction is based on a projected BCS (p-BCS) wave function:
\begin{equation}\label{pbcs}
|p{-}BCS \rangle = {\cal P} |BCS \rangle,
\end{equation}
where ${\cal P}$ is the projector onto the subspace of single occupied sites, 
and $|BCS \rangle$ is the ground state of
\begin{equation}\label{hambcs}
{\cal H}_{BCS} = -t\sum_{\langle R,R^\prime \rangle \sigma} 
c^\dag_{R,\sigma} c_{R^\prime,\sigma} 
+ \sum_{R,R^\prime} \Delta_{R,R^\prime} 
c^\dag_{R,\uparrow} c^\dag_{R^\prime,\downarrow} + H.c.,
\end{equation}
where $c^\dag_{R,\sigma}$ ($c_{R,\sigma}$) creates (destroys) an electron of
spin $\sigma$ at site $R=(x,y)$, $t=1$ is the nearest-neighbor hopping 
amplitude, and $\Delta_{R,R^\prime}=\Delta_{R^\prime,R}$ are real 
singlet pairing determining the symmetry of the BCS order parameter.
If the pairing involves only nearest neighbors sites along the coordinate 
directions, $\Delta^x$ and $\Delta^y$ (even under both reflections 
${\cal R}_x$ and ${\cal R}_y$), the BCS eigenstates can be labeled by further 
quantum numbers, i.e., $p_h=\pm 1$, associated to the particle-hole 
transformation
\begin{equation}\label{phole}
P_h c^{\dag}_{R,\sigma} P_h = {\rm sign}(\sigma) (-1)^{x+y} c_{R,-\sigma},
\end{equation}
which commutes with both ${\cal H}_{BCS}$ and ${\cal P}$, and the projected 
state~(\ref{pbcs}) has the same spatial symmetries of the spin Hamiltonian. 
However, the physical states that survive after projection are only those with 
$p_h=1$ for even $S^z$ or $p_h=-1$ for odd $S^z$. Therefore, all physical 
states have a given $P_h$ according to their $z$ component of the total spin.
In the BCS Hamiltonian, we have the freedom to take either
periodic boundary conditions (PBC), i.e., $c_{x,y} = c_{x+L,y}$, or 
antiperiodic ones (APBC), i.e., $c_{x,y}=-c_{x+L,y}$. Of course, in both cases,
after projection the wave function~(\ref{pbcs}) describes a spin state 
consistent with PBC. The lowest-energy state is obtained with PBC for 
$L=4 m+2$ and with APBC for $L=4 m$, while the lowest excitations correspond 
to the Gutzwiller projection of the BCS ground states with the other choice 
of the boundary conditions, namely APBC (PBC) for $L=4 m+2$ ($L=4 m$). 
In this case, the ground state of~(\ref{hambcs}) is degenerate because of
the presence of four zero-energy Bogoliubov modes, that can be identified 
as spinons, carrying spin $1/2$ and momenta $k_x=\pm \pi/2$.
With these objects, we can construct four zero-energy states 
with $p_h=1$ (that survive after projection): 
One triplet and three singlets. Indeed, two spinons can form 
$i)$ a triplet of momentum $k_x=\pi$ even under ${\cal R}_x$, 
$ii)$ one singlet with $k_x=\pi$ odd in ${\cal R}_x$, and 
$iii)$ one singlet with $k_x=0$ even under ${\cal R}_x$. A further singlet 
can be obtained by combining zero and four spinons, yielding 
$iv)$ a state with momentum $k_x=\pi$ and odd with respect to ${\cal R}_x$. 
All these states have ${\cal R}_y$ even. 
The two singlet states $ii)$ and $iv)$, belonging to the same symmetry
subspace, do not represent distinct excitations: In fact, numerical 
calculations show that their overlap {\it increases} with the size of the 
system. Analogously, the singlet $iii)$ is asymptotically identical 
to the ground state. Therefore, out of the four distinct states, we just 
obtain two independent excitations: A triplet and a singlet, precisely 
reproducing the lowest level of the ``tower of states'' predicted by conformal 
field theory in the non-frustrated case.~\cite{affleck}

\begin{table}
\caption{\label{tableI}
The low-energy states predicted by the p-BCS approach 
compared with the exact diagonalization data for $J_2/J_1=0.55$ on the
$10 \times 3$ ladder. The variational parameters are 
$\Delta^x=1.438$, $\Delta^y=-0.987$ and $\Delta^{xy}=0.457$.} 
\begin{ruledtabular}
\begin{tabular}{|c|c|c|c|c|c|c|c|}
Boundary & $k_x$ & $r_x$ & $r_y$ & Spin & $E_{vmc}$ & $E_0$ & Overlap \\
\hline
PBC  &   0   &  1 &  1 &  0 & -14.1281 & -14.2042 & 0.969 \\
\hline
APBC &   0   &  1 & -1 &  0 & -13.7997 & -13.9038 & 0.972 \\
\hline
APBC & $\pi$ &  1 &  1 &  1 & -13.7532 & -13.8689 & 0.923 \\
\hline
APBC & $\pi$ & -1 &  1 &  0 & -13.6404 & -13.8023 & 0.932 \\
\hline
APBC & $\pi$ &  1 & -1 &  0 & -13.0647 & -13.3265 & 0.882 \\
\end{tabular}
\end{ruledtabular}
\end{table}

Much more interesting is the case when the BCS Hamiltonian breaks some symmetry 
that is instead restored after projection. For instance, as already
emphasized in 2D,~\cite{rainbow} we can add a next-nearest neighbor 
pairing $\Delta^{xy}$, with odd reflection symmetry under 
${\cal R}_x$ and ${\cal R}_y$. In this case, both reflection and 
particle-hole symmetry (\ref{phole}) do not commute with ${\cal H}_{BCS}$ 
since the simultaneous presence of $\Delta^x$, $\Delta^y$, and $\Delta^{xy}$.
Instead, after projection onto the physical subspace with singly occupied 
sites, these symmetries are restored. Indeed, in this case, the BCS 
Hamiltonian is invariant under ${\tilde {\cal R}}_x=P_h \, {\cal R}_x$ 
and ${\tilde {\cal R}}_y= P_h \, {\cal R}_y$.
Therefore, the ground state of ${\cal H}_{BCS}$ has a well defined value 
of ${\tilde {\cal R}}_x$ and ${\tilde {\cal R}}_y$ and, after projection,
since $P_h$ is the identity in the physical Hilbert space, it has also well
defined values of the true reflection symmetries ${\cal R}_x$ and ${\cal R}_y$.
Since $P_h$ is no longer a symmetry of 
${\cal H}_{BCS}$, the eigenstates that were previously forbidden in the 
physical spectrum due to their wrong value of $P_h$ are instead now allowed. 
This approach predicts to a pair of new $r_y=-1$
excitations not present in the usual bosonization analysis: They are both 
singlets, even with respect to ${\cal R}_x$ and have different momenta: 
$v)$ $k_x=0$ and $vi)$ $k_x=\pi$. Remarkably, the first of these states has 
precisely the same quantum numbers of the lowest excitation emerging in the 
frustrated regime $J_2/J_1 \sim 0.55$, while the second one, although higher 
in energy, has a sudden drop in the same region, see Fig.~\ref{fig:spectrum}.

Further insight on the physics underlying this result comes from the structure
of the BCS excitations. The key feature induced by a non-vanishing 
$\Delta^{xy}$ coupling is the localization of the spinon wave function 
$\psi_j$ near the edges of the ladder (here $j$ labels the number of the leg): 
$\psi_{2j}=0$ and $\psi_{2j+1}=\tau^j$ with 
$\tau=-(t-2\Delta^{xy}+i\Delta^y)/(t+2\Delta^{xy}+i\Delta^y)$.
Only in the $\Delta^{xy}\to 0$ limit we get $\tau=-1$ and the spinon 
delocalizes among the chains. Instead, if $|\tau|\ne 1$, the gapless spinons 
become {\it edge states}.
This scenario is confirmed by a DMRG study on three- and five-leg ladder 
with {\it odd} number of sites per leg, so to have a free spinon trapped in 
the lattice.

\begin{figure}
\includegraphics[width=0.45\textwidth]{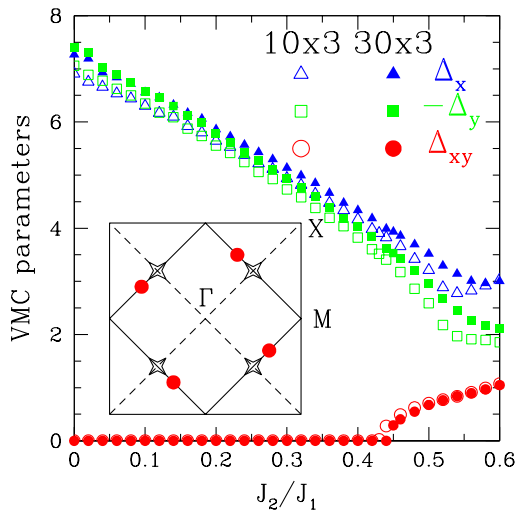}
\caption{\label{fig:phasediag}
(Color online) ``Phase diagram'' of the $J_1{-}J_2$ three-leg ladder found 
by the variational p-BCS state (see text). The Brillouin zone for the 2D
lattice is shown in the inset: Stars indicate 2D gapless spinon excitations 
for $\Delta^{xy}=0$ [at $(\pm \pi/2, \pm \pi/2)$], red dots for 
$\Delta^{xy} \ne 0$.} 
\end{figure}

After this detailed exam of the general properties of the p-BCS wave function, 
we show in Fig.~\ref{fig:phasediag} the ``phase diagram'' of the three-leg 
ladder, obtained by variational Monte Carlo. The optimal couplings 
$\Delta^x$, $\Delta^y$ and $\Delta^{xy}$ have been numerically determined. 
The most relevant feature is the stabilization of a non-vanishing 
$\Delta^{xy}$ for $J_2/J_1 \gtrsim 0.44$.
The variational estimate of the transition point is likely to be somewhat 
inaccurate, e.g., for one chain, spontaneous dimerization is found for
$J_2/J_1 \gtrsim 0.15$ instead of $J_2/J_1 \sim 0.24$.
Nevertheless we believe that the global trends emerging from this
phase diagram are representative of the real behavior of the model. 
Furthermore, we checked that dimer correlations do not display any clear 
tendency towards ordering. Note that, whenever the spinon spectrum is gapped
(e.g., when PBC are taken along the rungs), the same p-BCS wave function
describes a dimerized state for finite $n$.~\cite{chiral}

Finally, by considering the set of variational parameters obtained by 
Monte Carlo optimization of the ground-state wave function, we evaluated the 
elementary excitations on the $10 \times 3$ by a change in the boundary 
conditions along the chain, {\it without} a further optimization. 
In this way, beside the ground state, we obtained an explicit form of 
the excited states $i)$, $ii)$, $v)$, and $vi)$ discussed previously.
The variational energies and the overlaps with the exact lowest eigenstates 
are shown in Table~\ref{tableI} for $J_2/J_1=0.55$. The impressive accuracy 
of the p-BCS wave function for the full set of low-lying states provides a 
clear evidence in favor of the variational picture.

In conclusion, our results point toward the existence of gapless phases 
in the three-leg spin-1/2 $J_1-J_2$ Heisenberg antiferromagnet for all values 
of the next-nearest neighbor frustrating interaction. An unconventional 
gapless phase with an excitation spectrum characterized by the presence of 
low-energy edge states is stabilized upon frustration. Although we cannot 
exclude the possibility of an exponentially small dimerization 
(e.g., with a spin gap for three legs of order $10^{-8} J_1$), we believe that 
the new qualitative features that we found faithfully characterize the 
physics of the system. This homogeneous phase is accurately described by a 
projected BCS wave function representing an algebraic spin liquid. 
This scenario is even more plausible in 2D, especially considering that the 
exponentially small tail of the gap implied by the KT behavior should 
disappear, being replaced by a more conventional power-law with
$\Delta_{\infty} \sim (J_3-J_3^c)^\nu$, and that the same type of 
wave function remains accurate when increasing the (odd) number $n$ of legs. 
The strongly frustrated region in 2D is then a gapless state with 
incommensurate spinon excitations, that naturally result from the stabilization
of a finite $\Delta^{xy}$,~\cite{rainbow} as shown in Fig.~\ref{fig:phasediag}.
Such an exotic spin spectrum can be experimentally detected in 2D frustrated 
antiferromagnets, like ${\rm Li_2VOSiO_4}$ and 
${\rm VOMoO_4}$ under pressure.~\cite{carretta,carretta2}

We thank M. Fabrizio, A.A. Nersesyan, and O.A. Starykh for interesting 
discussions. This work has been partially supported by CNR-INFM 
and MIUR (COFIN 2004 and 2005).

\end{document}